\documentclass[twocolumn,english,aps,pra,superscriptaddress,showpacs,tightenlines]{revtex4-1}
\usepackage{amsmath}
\usepackage{graphicx}
\usepackage{amssymb}
\usepackage{CJK}
\usepackage{color}

\begin{document}

\begin{CJK*}{GBK}{song}

\title{Emergent Oscillating bound states in a semi-infinite linear waveguide with a point-like $\Lambda$-type quantum emitter driven by a classical field}
\author{Yuping \surname{He} }
\affiliation{Key Laboratory of Low-Dimension Quantum Structures and Quantum Control of Ministry of Education, Synergetic Innovation Center for Quantum Effects and Applications, Institute of Interdisciplinary Studies, Institute of Interdisciplinary Studies, Xiangjiang-Laboratory and Department of Physics, Hunan Normal University, Changsha 410081, China}
\author{Ge \surname{Sun} }
\affiliation{Key Laboratory of Low-Dimension Quantum Structures and Quantum Control of Ministry of Education, Synergetic Innovation Center for Quantum Effects and Applications, Institute of Interdisciplinary Studies, Institute of Interdisciplinary Studies, Xiangjiang-Laboratory and Department of Physics, Hunan Normal University, Changsha 410081, China}
\author{Jing \surname{Li}}
\affiliation{Key Laboratory of Low-Dimension Quantum Structures and Quantum Control of Ministry of Education, Synergetic Innovation Center for Quantum Effects and Applications, Institute of Interdisciplinary Studies, Institute of Interdisciplinary Studies, Xiangjiang-Laboratory and Department of Physics, Hunan Normal University, Changsha 410081, China}
\author{Ya \surname{Yang} }
\affiliation{Key Laboratory of Low-Dimension Quantum Structures and Quantum Control of Ministry of Education, Synergetic Innovation Center for Quantum Effects and Applications, Institute of Interdisciplinary Studies, Institute of Interdisciplinary Studies, Xiangjiang-Laboratory and Department of Physics, Hunan Normal University, Changsha 410081, China}
\author{Jing \surname{Lu} }
\affiliation{Key Laboratory of Low-Dimension Quantum Structures and Quantum Control of Ministry of Education, Synergetic Innovation Center for Quantum Effects and Applications, Institute of Interdisciplinary Studies, Institute of Interdisciplinary Studies, Xiangjiang-Laboratory and Department of Physics, Hunan Normal University, Changsha 410081, China}
\author{Lan \surname{Zhou}}
\thanks{Corresponding author}
\email{zhoulan@hunnu.edu.cn}
\affiliation{Key Laboratory of Low-Dimension Quantum Structures and Quantum Control of Ministry of Education, Synergetic Innovation Center for Quantum Effects and Applications, Institute of Interdisciplinary Studies, Institute of Interdisciplinary Studies, Xiangjiang-Laboratory and Department of Physics, Hunan Normal University, Changsha 410081, China}

\begin{abstract}
An oscillating bound state is a phenomenon where excitations mediated by the continuum modes oscillate persistently.
Although it is generated by the superposition of two bound states in the continuum (BICs), such phenomenon is said to
be unique to giant atoms. We present the phenomenon of an oscillating bound state with an alternative waveguide QED system,
which is a one-dimensional semi-infinite waveguide coupled to a \textit{point-like} quantum emitter. This \textit{point-like}
quantum emitter is $\Lambda$-type quantum system with one transition driven by a classical field.
\end{abstract}

\maketitle

\end{CJK*}\narrowtext

\section{Introduction}

The interaction between quantum emitters (QEs) and the quantized radiation
field can result in drastically different physical phenomena depending on
the detailed structure of the electromagnetic environment. In free space,
the QE-field coupling leads to an excited QE unavoidably decaying
towards its ground state by emitting a single photon to any of a continuum
of optical states each characterized by different wave vectors. Boundaries
and artificial dimensional reduction tailor the mode density of the
electromagnetic field, spontaneous emission is either enhanced, inhibited or
even completely hindered\cite{CookPRA35,Alberpra46,Meystr56,Zoller66}.
High-finesse cavities are exploited to strongly enhance the coupling of QEs
to one preferred mode and/or suppress the coupling to all unwanted modes,
then emitters and photons in such cavities exchange energy periodically.
Motivated by interfacing static qubits with flying qubits in quantum
networks, great attention has been paid on the physical platform called
waveguide QED systems\cite{RMP95(23)}. Waveguide QED refers to a scenario
where QEs interact with one dimensional (1D) bosonic modes of one
dimensional systems such as waveguides, optical fibers, microwave
transmission lines etc.. Since it allows encoding quantum information in
both photon and QE's states, many applications in quantum information
processing have been proposed, such as single-photon switches\cite%
{ShenPRL95,LanPRL101,ZhouOE30}, photon memory devices\cite%
{LanPRA78,gongPRA78,DongPRA79}, single-photon routers\cite%
{HoiPRL107,lanPRL111,LuPRA89,LuOE23,routPRA94,WeiPRA89,AhumPRA99,routPRR02},
and frequency converters\cite{LanPRA89,LiPRA104,YaAPL123,YaAPL124}. The
tightly focusing photons in 1D systems greatly enhance the atom-photon
interaction which gives rise to many intriguing phenomena, such as, strong
photon-photon quantum correlations\cite%
{ShenPRL98,LiaoPRA82,RoyPRL106,Zheng107,ShiPRA84,XuPRA90}, controllable
bi-photon bound states\cite{LanOE30,YaOE31}, controllable out-of-band
discrete levels\cite{ZhouPRA80,sunPRA100,WangPRA101,luOL49,LiNJP26},
radiations from a QE exponentially localized in the vicinity of the QE\cite%
{SEPRA89,SEPRA96}.

An hot topic is the study of QEs coupled to a semi-infinite linear
waveguide which is an infinite waveguide with one end behaving as a perfect
mirror\cite%
{ShenPRA87,PRA842011,TTCKPRA87,TTCKPRA90,SongCTP69,NMPRA104,PRA105Yi}. Since
some amount of radiation emitted by the QE is reflected back by the mirror to
the QE, a decay of the QE to its ground state is decelerated and
accelerated, or even inhibited to a nonzero constant for selected values of
the QE-mirror distance. The finite probability of excitations in the QE is
attributed to the emergence of a bound state in the continuum (BIC). A BIC
is general a localized eigenmode with energy eigenvalue residing directly in
the scattering continuum. Most works focus on the BIC emerged by a single
two-level emitter (2LE). The study on multiple BICs and their corresponding
phenomena --- oscillating bound states, is unique  to giant atoms. To the best
of our knowledge, fewer are discussed for a point-like QE coupled to a semi-infinite
linear waveguide. In this paper, we study the waveguide QED system made of a driven
three-level emitter (3LE) coupled to a semi-infinite waveguide with linear
dispertion relation, the point-like 3LE is also driven by a classical
field. By obtaining the delay-differential equation governing the time
evolution of the 3LE's excitation, the dynamics of spontaneous emission is
studied by varying the 3LE-mirror distance. Besides the similarity to the
behavior of the 2LE --- a long-lived population is characterized by a residual
steady value, the long time behavior of the 3LE is also characterized by
a stationary oscillation in its population. A deeper and more rigorous
insight into the long-lived population reveals that two BICs enable the
creation of persistently oscillating bound states. This result demonstrates
that the phenomenon of oscillating bound states is not unique to giant atoms
since the 3LE in our system is a point-like QE.

The paper is organized as follows. In Sec.~\ref{Sec:2}, we present a
physical model of a 1D semi-infinite waveguide coupled to a point-like 3LE
with a $\Lambda $-type configuration. In Sec.~\ref{Sec:3}, after obtaining
the delay-differential equation for the 3LE's amplitudes in the
one-excitation subspace, we study the time evolution of the 3LE's excitation
when the 3LE is initial in its excited state and the waveguide is initially
in the vacuum state. Some typical examples of the dynamics are shown
according to the relation between the 3LS-mirror distance and the coherence
length. In Sec.~\ref{Sec:4}, the persistently oscillation in the long-lived
dynamic is discussed, it is a phenomenon of an oscillating bound state which
is a result of coexisting bound modes. In Sec.~\ref{Sec:5}, the total
intensity emitted by the 3LE is studied, the trapping of the excitation and
the interference of the two localized field modes are found. Then we make our
conclusion in Sec.~\ref{Sec:6}.


\section{\label{Sec:2}Model setup}

\begin{figure}[tbph]
\includegraphics[width=8 cm,clip]{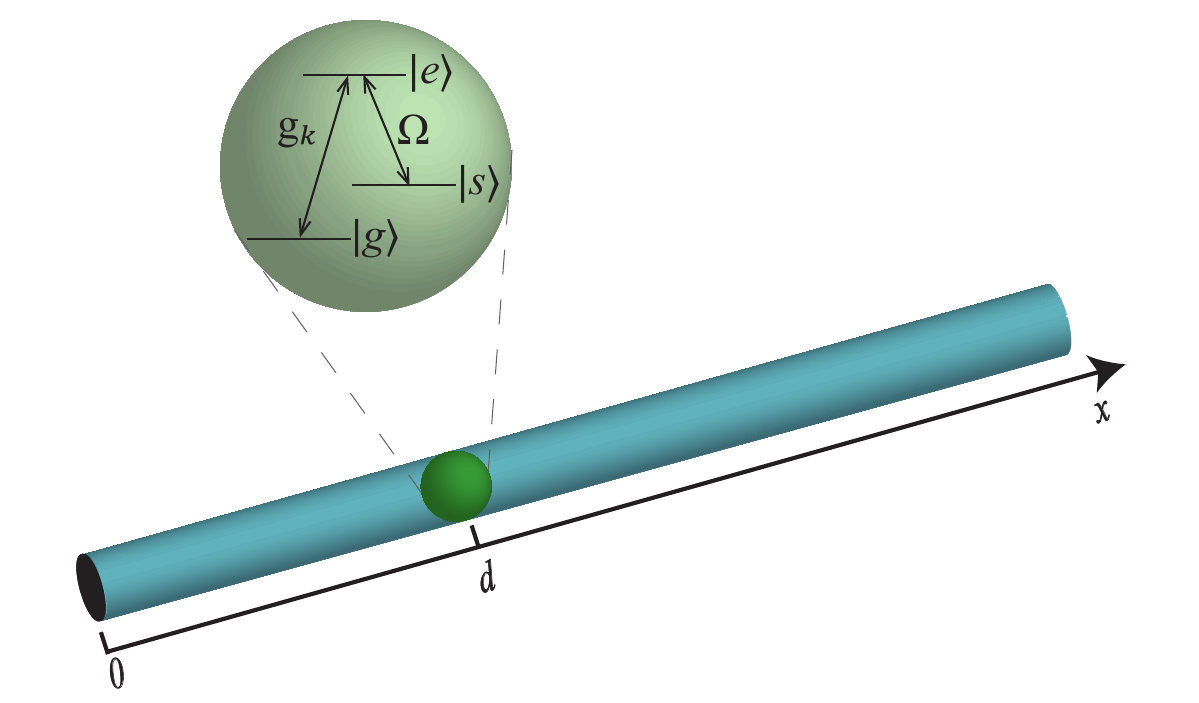}
\caption{(color online)Sketch of the setup: a three-level system located at $%
x=d$ coupled to a terminated waveguide whose termination is a perfect mirror
at $x=0$.}
\label{fig1}
\end{figure}
As illustrated in Fig.~\ref{fig1}, a single 3LE with a lower state $%
|g\rangle $, a metastable state $|s\rangle $ and an excited state $|e\rangle
$ is located at position $d$ along the $x$ axis, its energy of the states $%
i=g,s,e$ is denoted as $\omega _{g},\tilde{\omega}_{s},\omega _{e}$. One
transition $|e\rangle \leftrightarrow |s\rangle $ is driven by an external
classical field of frequency $\omega _{l}$ with the Rabi frequency $\Omega $%
, and the other transition $|e\rangle \leftrightarrow |g\rangle $ is dipole
coupled to the EM field modes of a 1D semi-infinite waveguide along the $x$
axis with coupling strengths $g_{k}=\sqrt{\Gamma \upsilon /\pi }\sin \left(
kd\right) $, where $\upsilon $ is the speed of light in the waveguide. The
end of the waveguide at $x=0$ behaves as a perfect mirror. Within the
rotating-wave approximation, the Hamiltonian $\hat{H}=\hat{H}_{A}+\hat{H}_{F}+\hat{H}_{I}$ in the
rotating frame is split into three pieces: the interaction between the 3LE
and the waveguide modes
\begin{equation}
\hat{H}_{I}=\int_{-\infty }^{+\infty }dkg_{k}\left( \hat{\sigma}_{ge}\hat{a}%
_{k}^{\dag }+\hat{\sigma}_{eg}\hat{a}_{k}\right)  \label{1-01}
\end{equation}%
is written in the electric dipole approximation, the Hamiltonian for the 3LE%
\begin{equation}
\hat{H}_{A}=\omega _{e}\hat{\sigma}_{ee}+\omega _{s}\hat{\sigma}_{ss}+\Omega
\left( \hat{\sigma}_{se}+\hat{\sigma}_{es}\right)  \label{1-02}
\end{equation}%
and the free Hamiltonian for the guided modes of the waveguide%
\begin{equation}
\hat{H}_{F}=\int dk\omega _{k}\hat{a}_{k}^{\dag }\hat{a}_{k}.  \label{1-03}
\end{equation}%
Here, the energy of the 3LE ground state $\omega _{g}=0$ and $\omega _{s}=%
\tilde{\omega}_{s}+\omega _{l}$, $\hat{\sigma}_{ij}=|i\rangle \left\langle
j\right\vert $ is the operator related to the 3LE, the annihilation
(creation) operator $\hat{a}_{k}$($\hat{a}_{k}^{\dag }$) annihilates
(creates) a photon in the mode $k$ with wavenumber $k$ satisfying the dispersion
relation $\omega _{k}=\upsilon k$, and they obey the bosonic commutation rule
$\left[ \hat{a}_{k},\hat{a}_{k^{\prime }}^{\dag }\right] =\delta \left( k-k^{\prime }\right) $.
The total Hamiltonian conserves the number of excitations so that the state
of the QE-field system at time $t$ reads
\begin{equation}
\left\vert \Psi \left( t\right) \right\rangle =C_{e}\left( t\right)
\left\vert e0\right\rangle +C_{s}\left( t\right) \left\vert s0\right\rangle
+\int dk\beta _{k}\left( t\right) \left\vert g1_{k}\right\rangle
\label{1-04}
\end{equation}%
in the single-excitation subspace, where $C_{e}\left( t\right) $/$%
C_{s}\left( t\right) $ is the amplitude to find the 3LE in state $\left\vert
e\right\rangle $/$\left\vert s\right\rangle $ and the photon in the vacuum
state $\left\vert 0\right\rangle $ of the waveguide field, and $\beta
_{k}\left( t\right) $ is the amplitude to find the 3LE in its ground state
and exactly one photon in mode $k$. The Schr\"{o}dinger equation transforms $%
\left\vert \Psi \left( t\right) \right\rangle $ into three coupled equaitons
of motion for the 3LE and field excitation amplitudes
\begin{subequations}
\label{1-05}
\begin{align}
\dot{C}_{e}\left( t\right) & =-\mathrm{i}\omega _{e}C_{e}\left( t\right) -%
\mathrm{i}\Omega C_{s}\left( t\right) -\mathrm{i}\int dkg_{k}\beta
_{k}\left( t\right) , \\
\dot{C}_{s}\left( t\right) & =-\mathrm{i}\omega _{s}C_{s}\left( t\right) -%
\mathrm{i}\Omega C_{e}\left( t\right) \\
\dot{\beta}_{k}\left( t\right) & =-\mathrm{i}\omega _{k}\beta _{k}\left(
t\right) -\mathrm{i}g_{k}C_{e}\left( t\right) .
\end{align}


\section{\label{Sec:3}Spontaneous Emission Dynamics}


In this section, we consider an initially excited 3LE in the absence of any
photon, i.e., $\left\vert \Psi \left( 0\right) \right\rangle =C_{e}\left(
0\right) \left\vert e0\right\rangle +C_{s}\left( 0\right) \left\vert
s0\right\rangle $. The photonic degree of freedom can be completely
eliminated by formally integrating the equation for $\beta _{k}\left(
t\right) $ and inserting it into the former two equations in Eq.(\ref{1-05}%
), then the delay-differential equation for the 3LE's excitation reads
\end{subequations}
\begin{subequations}
\label{2-00}
\begin{eqnarray}
\dot{C}_{s}\left( t\right) &=&-\mathrm{i}\omega _{s}C_{s}\left( t\right) -%
\mathrm{i}\Omega C_{e}\left( t\right)  ,\\
\dot{C}_{e}\left( t\right) &=&-\left( i\omega _{e}+\frac{\Gamma }{2}\right)
C_{e}\left( t\right) -\mathrm{i}\Omega C_{s}\left( t\right)  \notag \\
&&+\frac{\Gamma }{2}C_{e}\left( t-\tau \right) \Theta \left( t-\tau \right).
\end{eqnarray}%
Here, $\Theta \left( t\right) $ is the Heaviside step function. By further
introducing $C_{e}\left( t\right) =c_{e}\left( t\right) e^{-\mathrm{i}\omega
_{e}t},C_{s}\left( t\right) =c_{s}\left( t\right) e^{-\mathrm{i}\omega
_{e}t} $, the delay-differential equation becomes
\end{subequations}
\begin{subequations}
\label{2-01}
\begin{eqnarray}
\dot{c}_{s}\left( t\right) &=&\mathrm{i}\delta c_{s}\left( t\right) -\mathrm{%
i}\Omega c_{e}\left( t\right) , \\
\dot{c}_{e}\left( t\right) &=&-\frac{\Gamma }{2}c_{e}\left( t\right) -%
\mathrm{i}\Omega c_{s}\left( t\right) \notag \\
&&+\frac{\Gamma }{2}e^{\mathrm{i}\omega _{e}\tau }c_{e}\left( t-\tau \right)
\Theta \left( t-\tau \right).
\end{eqnarray}%
with the detuning $\delta =\omega _{e}-\omega _{s}$. There are three
distinct processes that contribute to the spontaneous emission of the
excited state in Eq.(\ref{2-01}): (1) the 3LE relaxing to its ground state
and emitting a photon in the waveguide characterized by the spontaneous
decay rate $\Gamma /2$; (2) a coherent energy exchange between the states $%
|e\rangle $ and $|s\rangle $ because of the presence of the driving field;
(3) the effect of the retarded radiation on the 3LE due to a delay time $%
\tau =2d/\upsilon $ that the light emitted from the 3LE needs for a distance
3LE-mirror-3LE. When $\Omega =0$ and $\tau \rightarrow \infty $, eq.(\ref%
{2-01}b) leads to the usual exponential decay with rate $\Gamma $ of the 2LE
to its ground state accompanied by an irreversible release of energy to the
vacuum of a 1D infinite waveguide. When $\Omega =0$, the metastable state
involves freely, eq.(\ref{2-01}b) depicts the spontaneous dynamics of a 2LE
interacting with a 1D semi-infinite waveguide\cite%
{DongPRA79,PRA842011,TTCKPRA87,TTCKPRA90,SongCTP69}. The QE's excitation
is determined by the ratio of the characteristic wavelength $\lambda
_{e}=2\pi v/\omega _{e}$ and the coherence length $L_{c}=v/\Gamma $. When $\tau
\rightarrow 0$, two dressed states
\end{subequations}
\begin{subequations}
\label{2-02}
\begin{align}
\left\vert +\right\rangle & =\frac{\Omega }{\sqrt{2\Delta \left( \Delta
-\delta /2\right) }}\left\vert e\right\rangle +\sqrt{\frac{\Delta -\delta /2%
}{2\Delta }}\left\vert s\right\rangle \\
\left\vert -\right\rangle & =\frac{\Omega }{\sqrt{2\Delta \left( \Delta
+\delta /2\right) }}\left\vert e\right\rangle -\sqrt{\frac{\Delta +\delta /2%
}{2\Delta }}\left\vert s\right\rangle
\end{align}%
of the corresponding energies $\omega _{\pm }=\bar{\omega}\pm \Delta $ are
formed with $\bar{\omega}=\left( \omega _{e}+\omega _{s}\right) /2$, so the
3LE undergoes an oscillation between $|e\rangle $ and $|s\rangle $ with the
frequency $\Delta =\sqrt{\delta ^{2}/4+\Omega ^{2}}$. When $\tau \rightarrow
\infty $, eq.(\ref{2-01}) depicts the spontaneous dynamics of a 3LE
interacting with a 1D infinite waveguide, and we have plotted the time
variation of the probability $P_{e}=\left\vert C_{e}\left( t\right)
\right\vert ^{2}$ in Fig.\ref{fig2} with the intial condition $C_{e}\left(
0\right) =1,C_{s}\left( 0\right) =0$. It can be found that when $\delta =0$,
the 3LE's excitation $P_{e}$ decays faster than that of 2LE in the regime $%
\Omega \in \left( 0,\Gamma /4\right] $, so the coupling to an another level
increases the loss of the energy from its initial state in addition to its
releasing the energy to the vacuum; the atomic excitation $P_{e}$ exhibits
decaying oscillations in the regime $\Omega >\Gamma /4$, and the decaying
oscillations gradually disappear with the increasing of the detuning.
\begin{figure}[tbph]
\includegraphics[width=8 cm]{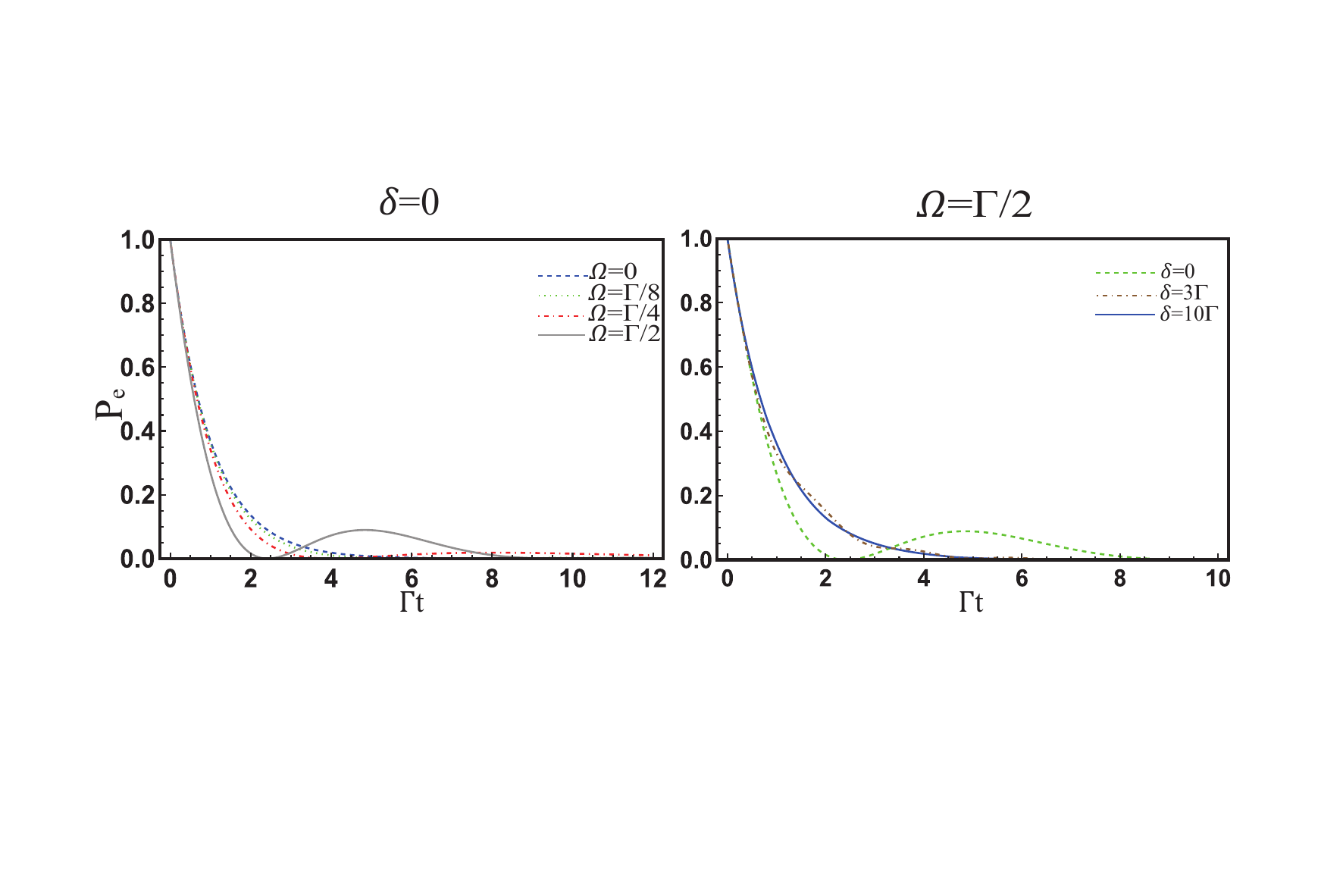}
\caption{(color online) Atomic excitation $P_{e}$ as a function of time for
(a) $\protect\delta =0$ and (b) $\Omega =\Gamma /2$ when $\protect\tau %
\rightarrow \infty $.}
\label{fig2}
\end{figure}
As a result, the energies are all carried by the spontaneously emitted wave
and leak into the waveguide.
\begin{figure}[tbph]
\includegraphics[width=8 cm]{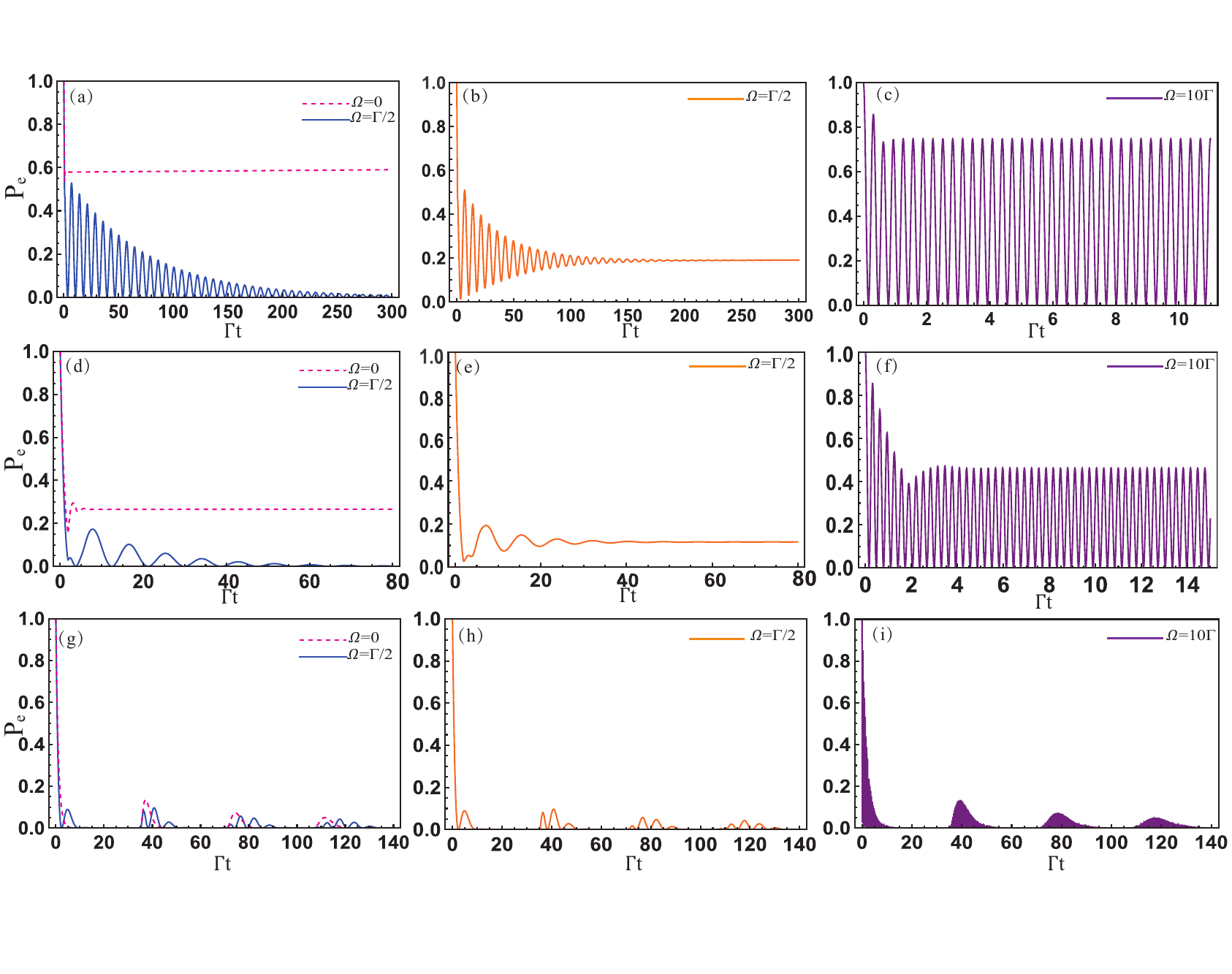}
\caption{(color online) The dynamics of atomic excitation $P_{e}$ in three
different regimes when $\protect\delta =0$: upper panels $d<L_{c}$, (a) $%
d=0.316L_{c}$, (b) $d=0.315L_{c}$, (c) $d=0.316L_{c}$, middle panels $d\sim
L_{c}$,(d-f) $d=0.94L_{c}$, lower panels $d\gg L_{c}$, (g-i) $d=18L_{c}$.}
\label{fig3}
\end{figure}

When $\tau $ is finite, the emitted photon reflected by the boundary can be
absorbed or scattered by the 3LE, giving rise to the transmitted and
reflected waves. The reflected waves return to the boundary. Since the
atomic reabsorption and scattering of the emitted photon can occur at times $%
t=n\tau $, the process would be repeated, leading to the inteference between
the light emitted in the past and the radiation in the present. The
interference behavior is even more richer for a 3LE since multiple
transition frequencies are involved. So in the following discussion,
attention is paid on the case with $\Omega >\Gamma /4$ and $\delta =0$. Fig.%
\ref{fig3} plotted by numerically solving Eq.(\ref{2-01}) shows the dynamics
of 3LE's excitation $P_{e}$ in three different regimes with the intial
condition $C_{e}\left( 0\right) =1,C_{s}\left( 0\right) =0$: the 3LE close
to the boundary characterized by $d<L_{c}$ (see Fig.\ref{fig3}a-c),
3LE-mirror separation comparable to the coherence length $d\sim L_{c}$ (see
Fig.\ref{fig3}d-f), and a large 3LE-mirror spacing characterized by $d\gg
L_{c}$ (see Fig.\ref{fig3}g-i). The time evolution of atomic excitation with
$\Omega =0$ is displayed for comparison in Fig.\ref{fig3}a,d,g. To maintain
the relation between $d$ and $L_{c}$, we have chosen $\Omega =10\Gamma $ in
Fig.\ref{fig3}c,f,i. It can be found from Fig.\ref{fig3}a that destructive
interference occurring at $d=n\lambda _{e}/2$ leads to a partial trap of the
QE's excitation due to the emergence of the QE-photon bound state for a 2LE,
the 3LS is driven toward underdamped and relaxes to its ground state in the
large-time limit. When $d=n\lambda _{+}/2$ ($n\lambda _{-}/2$) with $\lambda
_{\pm }=2\pi v/\omega _{\pm }$, the long-lived QE's excitation can also be
observed in Fig.\ref{fig3}b. However, there is a stationary oscillation in
Fig.\ref{fig3}c with $d=n\lambda _{o}/2$ and the oscillating wavelength $%
\lambda _{o}=2\pi v/\Omega $. As the 3LS-mirror separation increases, the
retardation effects of the waveguide field become relevant. When the width
of the wavepacket of the emitted photon is comparable to the 3LE-mirror
separation, it is possible for the emitted photon reflected by the mirror to
interfere with the wave emitted by the 3LE, so one can observe the phenomen
similar to that of regime $d<L_{c}$ but with a lower probability in Fig.\ref%
{fig3}e,f. As the 3LE-mirror separation increases further so that $d\gg
L_{c} $, interference plays no role (see Fig.\ref{fig3}g-i), the 3LE
displays the same behavior as a 3LE inside an infinite waveguide at the
beginning. After time $t>\tau $, the emitted photon is bounced back and
forth between the 3LE and the mirror until its amplitude is damped to zero,
and a broader shape is acquired by each bounce.

To understand why the 3LE holds a significant amount of excitation, we write
the state of the 3LE-field system at time $t$ in terms of dressed states
\end{subequations}
\begin{eqnarray*}
\left\vert \Psi \left( t\right) \right\rangle &=&c_{+}\left( t\right) e^{-%
\mathrm{i}\omega _{+}t}\left\vert +0\right\rangle +c_{-}\left( t\right) e^{-%
\mathrm{i}\omega _{-}t}\left\vert -0\right\rangle \\
&&+\int dk\beta _{k}\left( t\right) \left\vert g1_{k}\right\rangle ,
\end{eqnarray*}%
and the delay-differential equation becomes
\begin{subequations}
\label{2-03}
\begin{eqnarray}
\dot{c}_{+}\left( t\right) &=&\frac{\Gamma \left( \Delta +\delta /2\right) }{%
4\Delta }\left[ e^{\mathrm{i}\omega _{+}\tau }c_{+}\left( t-\tau \right)
\Theta \left( t-\tau \right) -c_{+}\left( t\right) \right]  \nonumber \\
&&+\frac{\Omega \Gamma }{4\Delta }e^{\mathrm{i}2\Delta t}\left[ e^{\mathrm{i}%
\omega _{-}\tau }c_{-}\left( t-\tau \right) \Theta \left( t-\tau \right)
-c_{-}\left( t\right) \right] \nonumber \\ \\
\dot{c}_{-}\left( t\right) &=&\frac{\Gamma \left( \Delta -\delta /2\right) }{%
4\Delta }\left[ c_{-}\left( t-\tau \right) e^{\mathrm{i}\omega _{-}\tau
}\Theta \left( t-\tau \right) -c_{-}\left( t\right) \right]  \nonumber \\
&&+\frac{\Omega \Gamma }{4\Delta }e^{-\mathrm{i}2\Delta t}\left[ e^{\mathrm{i%
}\omega _{+}\tau }c_{+}\left( t-\tau \right) \Theta \left( t-\tau \right)
-c_{+}\left( t\right) \right] \nonumber \\
\end{eqnarray}%
In the regime $d<L_{c}$, the phases radically modify the behavior of the
spontaneous emission\cite{ShenPRA87,SongCTP69,HuQIP19},the phenomenon can be
understood by letting $\tau \rightarrow 0$ in Eq.(\ref{2-03})
\end{subequations}
\begin{eqnarray*}
\dot{c}_{+}\left( t\right) &=&\frac{\Gamma \left( \Delta +\delta /2\right) }{%
4\Delta }\left( e^{\mathrm{i}\omega _{+}\tau }-1\right) c_{+}\left( t\right)
\\
&&+\frac{\Omega \Gamma }{4\Delta }e^{\mathrm{i}2\Delta t}\left( e^{\mathrm{i}%
\omega _{-}\tau }-1\right) c_{-}\left( t\right) \\
\dot{c}_{-}\left( t\right) &=&\frac{\Gamma \left( \Delta -\delta /2\right) }{%
4\Delta }\left( e^{\mathrm{i}\omega _{-}\tau }-1\right) c_{-}\left( t\right)
\\
&&+\frac{\Omega \Gamma }{4\Delta }e^{-\mathrm{i}2\Delta t}\left( e^{\mathrm{i%
}\omega _{+}\tau }-1\right) c_{+}\left( t\right)
\end{eqnarray*}%
Obviously, the phases $\omega _{\pm }\tau $ determine whether the delayed
amplitude leads to an accelerated or decelerated decay. When one phase (say $%
\omega _{+}\tau $) satifies $\omega _{+}\tau =2n\pi $ but the other doesnot
(i.e., $\omega _{-}\tau \neq 2n\pi $), the decay of upper dressed state $%
\left\vert +\right\rangle $ is greatly inhibited, and it only exchanges
energy with the lower dressed state $\left\vert -\right\rangle $, the
probability of state $\left\vert -\right\rangle $ decays in an exponential
function at rate $\Gamma \left( \Delta -\delta /2\right) \left( \cos \omega
_{-}\tau -1\right) /\left( 4\Delta \right) $, and becomes zero after a
sufficiently long time, however, a significant amount of excitation is still
held in the state $\left\vert +\right\rangle $, it is why one can observed a
steady value of population in Fig.\ref{fig3}b.\ Moreover, it is
possible to find two integers $n_{\pm }$ satifying $\omega _{\pm }\tau =$ $%
2n_{\pm }\pi $ simultaneously, so destructive interference inhibits emission
into guided modes, and two waves of frequency $\omega _{\pm }$ are
generated, their interference displays a net resulting wave of average
frequency $\bar{\omega}$ which oscillates in strength with a frequency $%
\left\vert \omega _{+}-\omega _{-}\right\vert /2=\Delta $.


\section{\label{Sec:4}Bound states in the continuum}

\begin{figure}[tbph]
\includegraphics[width=8 cm]{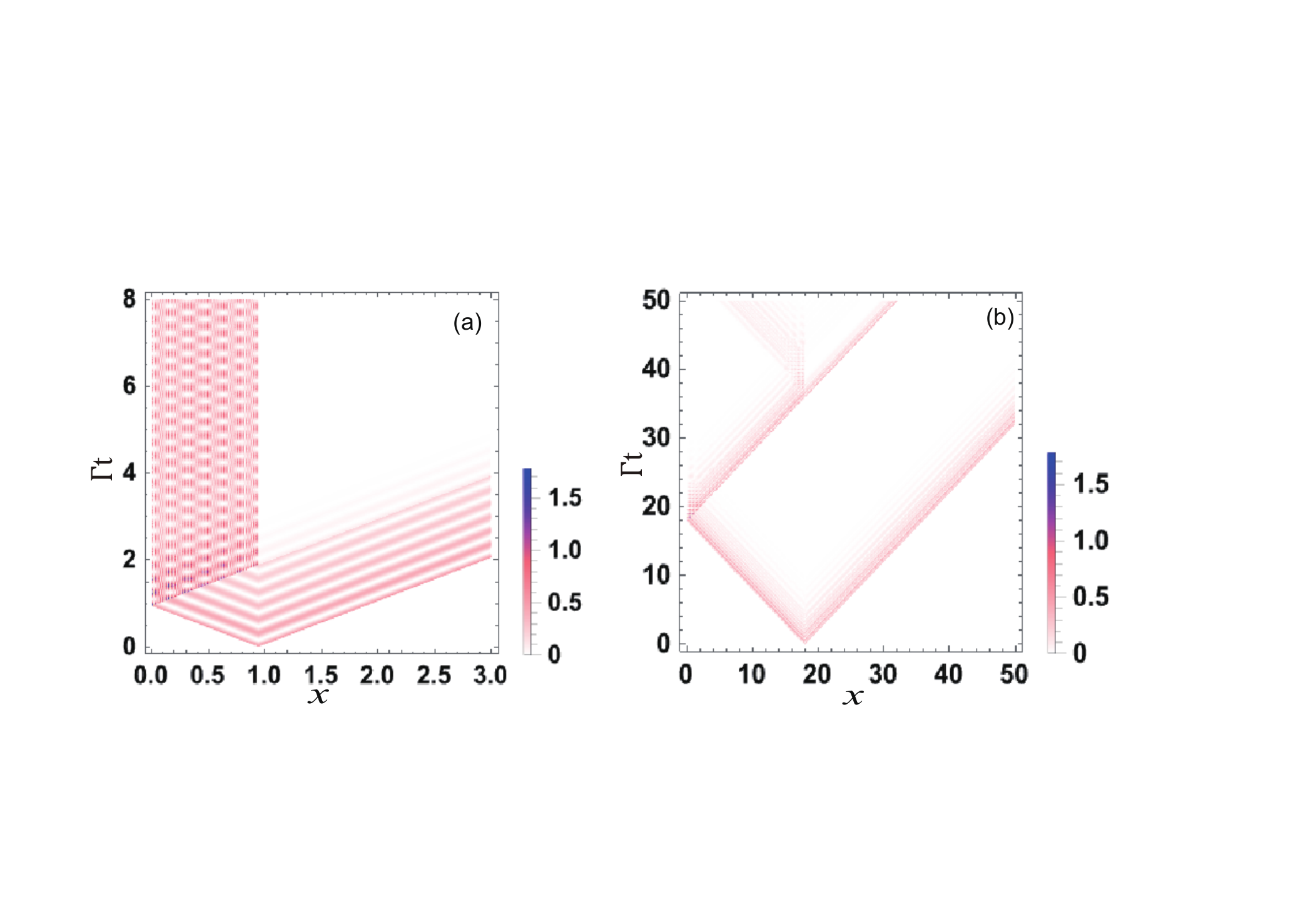}
\caption{(color online) the total intensity $I\left( x,t\right) $ in the $x-t$
parameter plane for $d\sim L_{c}$ (a) and $d\gg L_{c}$ (b) with the same parameters
as in Fig.~\ref{fig3}(f) and ~\ref{fig3}(i) respectively. }
\label{fig4}
\end{figure}

Previous studies\cite{LanPRL101,RistPRA78,LanPRA78,gongPRA78} have shown that a
single atom in a waveguide can act as a tunable mirror. In this
perspective, the 1D confined space sandwiched between the atom and the
mirror can be considered as a cavity\cite{DongPRA79}. A perfect cavity
emerges in this region when the cavity photon mode of energy resonates with
the level of the atomic excited state. We show in this section that energies
$\omega _{\pm }$ are the eigenvalues of the system, whose corresponding
states are BICs. The existence of oscillating BICs gives rise to the
persistent oscillation in Fig.\ref{fig3}i. Performing the Laplace
transformation on Eq.(\ref{2-00})
\begin{subequations}
\label{3-01}
\begin{eqnarray}
C_{e}\left( p\right) &=&\frac{C_{e}\left( 0\right) \left( p+\mathrm{i}\omega
_{s}\right) -\mathrm{i}\Omega C_{s}\left( 0\right) }{\left( p+\mathrm{i}%
\omega _{e}+\frac{\Gamma }{2}-\frac{\Gamma }{2}e^{-p\tau }\right) \left( p+%
\mathrm{i}\omega _{s}\right) +\Omega ^{2}} \nonumber \\ \\
C_{s}\left( p\right) &=&\frac{\left( p+\mathrm{i}\omega _{e}+\frac{\Gamma }{2%
}-\frac{\Gamma }{2}e^{-p\tau }\right) C_{s}\left( 0\right) -\mathrm{i}\Omega
C_{e}\left( 0\right) }{\left( p+\mathrm{i}\omega _{e}+\frac{\Gamma }{2}-%
\frac{\Gamma }{2}e^{-p\tau }\right) \left( p+\mathrm{i}\omega _{s}\right)
+\Omega ^{2}} \nonumber \\
\end{eqnarray}%
gives the solution of the amplitudes, but most importantly, each pole is
associated with a (generally unstable) state of the system, and its nonzero
real component leads to the decay of the 3LS. A long-lived state is
presented by an exponential factor in the numerator like $e^{-\mathrm{i}%
\omega t}$, so we set $p=-\mathrm{i}\omega $, and the roots of the
denominator in Eq.(\ref{3-01}) on the imaginary axis require
\end{subequations}
\begin{subequations}
\label{3-02}
\begin{eqnarray}
0 &=&\frac{\Gamma }{2}\left( 1-\cos \omega \tau \right) \left( \omega
-\omega _{s}\right) \\
0 &=&\left( \omega -\omega _{e}\right) \left( \omega -\omega _{s}\right)
-\Omega ^{2}
\end{eqnarray}%
to be satisfied. Then we obtain $\omega \tau =2n\pi ,n\in N$ and $\omega
=\omega _{\pm }$, i.e., the conditions for the existence of the BIC. It is
possible to find two integers $n_{\pm }$ satisfying Eq. (\ref{3-02})
simultaneously. After all unstable state die out, the amplitude for the
excited state reads

\end{subequations}
\begin{eqnarray}
C_{e}\left( t\right) &=&\frac{e^{-\mathrm{i}\omega _{+}t}2\cos ^{2}\theta }{%
\tau \Gamma \cos ^{2}\theta +2}\frac{\left( \delta +2\Delta \right)
C_{e}\left( 0\right) -2\Omega C_{s}\left( 0\right) }{\delta +2\Delta }
\nonumber \\
&&+\frac{e^{-\mathrm{i}\omega _{-}t}2\sin ^{2}\theta }{\tau \Gamma \sin
^{2}\theta +2}\frac{\left( \delta -2\Delta \right) C_{e}\left( 0\right)
-2\Omega C_{s}\left( 0\right) }{\delta -2\Delta }\nonumber \\  \label{3-03}
\end{eqnarray}%
where $\sin \theta =\sqrt{\frac{2\Delta -\delta }{4\Delta }}$and $\cos
\theta =\sqrt{\frac{2\Delta +\delta }{4\Delta }}$. Obviously, residues in
two BICs are possible if they exist, so the 3LE's excitation oscillates
persistently with frequency $\Delta $. For the convenience of the latter
discussion, we introduce
\begin{subequations}
\label{3-04}
\begin{eqnarray}
C_{e}^{+}\left( t\right) &=&\frac{e^{-\mathrm{i}\omega _{+}t}2\cos
^{2}\theta }{\tau \Gamma \cos ^{2}\theta +2}\frac{\left( \delta +2\Delta
\right) C_{e}\left( 0\right) -2\Omega C_{s}\left( 0\right) }{\delta +2\Delta
} \nonumber \\ \\
C_{e}^{-}\left( t\right) &=&\frac{e^{-\mathrm{i}\omega _{-}t}2\sin
^{2}\theta }{\tau \Gamma \sin ^{2}\theta +2}\frac{\left( \delta -2\Delta
\right) C_{e}\left( 0\right) -2\Omega C_{s}\left( 0\right) }{\delta -2\Delta
} \nonumber \\
\end{eqnarray}

\section{\label{Sec:5}Output Field}

The annihilation operator of the waveguide field at position $x$ can be
given in the mode expansion as
\end{subequations}
\begin{equation}
\hat{E}\left( x\right) =\sqrt{\frac{2}{\pi }}\int dk\hat{a}_{k}\sin \left(
kx\right) .  \label{4-01}
\end{equation}%
the total intensity emitted by the 3LE $I\left( x,t\right) =\left\langle
\Psi \left( t\right) \right\vert \hat{E}^{\dag }\left( x\right) \hat{E}%
\left( x\right) \left\vert \Psi \left( t\right) \right\rangle \equiv
\left\vert \psi \left( x,t\right) \right\vert ^{2}$ is usually measured by
the local field profile $\psi \left( x,t\right) $. For a photon detector
lying at position $x$, we obtain%
\begin{eqnarray}
\psi \left( x,t\right)  &=&\sqrt{\frac{\Gamma }{2\upsilon }}C_{e}\left(
t_{x}^{-}-\frac{\tau }{2}\right) \Theta \left( t_{x}^{-}-\frac{\tau }{2}%
\right)   \label{4-02} \\
&&-\sqrt{\frac{\Gamma }{2\upsilon }}C_{e}\left( t_{x}^{+}-\frac{\tau }{2}%
\right) \Theta \left( t_{x}^{+}-\frac{\tau }{2}\right) \Theta \left(
d-x\right)   \notag \\
&&-\sqrt{\frac{\Gamma }{2\upsilon }}C_{e}\left( t_{x}^{-}+\frac{\tau }{2}%
\right) \Theta \left( t_{x}^{-}+\frac{\tau }{2}\right) \Theta \left(
x-d\right)   \notag
\end{eqnarray}%
with $t_{x}^{\pm }=t\pm \frac{x}{\upsilon }$. In Fig.~\ref{fig4}, we plot
the time evolution of the total intensity $I\left( x,t\right) $ in the $x-t$
parameter plane for (a) $d\sim L_{c}$ and (b) $d\gg L_{c}$. One can observe
that a right-going wave propagates far away from the 3LE, a left-going wave
propagates toward the mirror and bounces back by the mirror in Fig.~\ref{fig4}.
However, a ray trajectory and the intensity damping can be clearly seen in
Fig.~\ref{fig4}b. Fig.~\ref{fig4}(a) presents a constructive and destructive
of waves in the regime between the 3LE and the mirror, and its interference
pattern is unchanged in the long time limit.

If the condition for the existence of the two BICs are satisfied, the
excite-state population does not fully decay to its ground state, the
residual field intensity mainly comes from two steady states at the long
time limit, so we write $\psi \left( x,t\right) $ as%
\begin{equation}
\psi \left( x,t\right) =\psi _{+}\left( x,t\right) +\psi _{-}\left(
x,t\right)   \label{4-03}
\end{equation}%
a superposition of two field profile $\psi _{\pm }\left( x,t\right) $ at
position $x$ of two BICs with%
\begin{eqnarray}
\psi _{\pm }\left( x,t\right)  &=&\sqrt{\frac{\Gamma }{2\upsilon }}%
C_{e}^{\pm }\left( t-\frac{x+d}{\upsilon }\right)   \label{4-04} \\
&&-\sqrt{\frac{\Gamma }{2\upsilon }}C_{e}^{\pm }\left( t+\frac{x-d}{\upsilon
}\right) \Theta \left( d-x\right)   \notag \\
&&-\sqrt{\frac{\Gamma }{2\upsilon }}C_{e}^{\pm }\left( t-\frac{x-d}{\upsilon
}\right) \Theta \left( x-d\right)   \notag
\end{eqnarray}%
at time $t>\frac{x+d}{\upsilon }$. Here, each BIC corresponds to a path and
the coherence property is obviously reflected by the intensity interference
of two field profiles. For a detector placed at $x$, we obtain
\begin{equation}
\psi _{\pm }\left( x,t\right) =\left\{
\begin{array}{c}
\mathrm{i}\sqrt{\frac{2\Gamma }{\upsilon }}C_{e}^{\pm }\left( 0\right) \sin
\left( \omega _{\pm }\frac{x-d}{\upsilon }\right) e^{-\mathrm{i}\omega _{\pm
}t}\text{ for }x<d \\
0\text{ for }x\geq d%
\end{array}%
\right. .  \label{4-05}
\end{equation}%
Obviously, the field is confined to the region sandwiched between the atom and
the mirror, and forms two perfectly sinusoidal waves. The expression in Eq.(\ref%
{4-03}) indicates that the oscillation result from the interference of the
modes with different frequencies originated from the dressed states of the
3LE at different frequencies.


\section{\label{Sec:6}Conclusion}


We have studied the emission process of a point-like $\Lambda $-type 3LS
with one electric-dipole coupled to the modes of a 1D single-end waveguide
and the other electric-dipole coupled to a classical field. The time
evolution of the 3LS's excited state are distance-dependent due to the
interference between the light emitted in the past and the radiation in the
present. Despite its nonexponential decay which can exhibit oscillations for
the 3LS-mirror distance larger than the coherent length, three long-time
behaviors for the 3LS-mirror distance smaller than the coherent length are
shown: dampened to zero, dampened to a steady value, and dampened to a
persistent oscillation. A steady residue represents a clear signature of
single-photon trapping. This amount of trapped excitation owes to the
formation of a bound state appearing inside the continuous energy spectrum,
which is well-known in a semi-infinite waveguide coupling to a point-like
2LE. The novel feature of a point-like 3LE in waveguide QED is the
persistent oscillation, which is a result of coexisting BICs. The field
profiles of the BICs are two perfectly sinusoidal waves confined in the
region between the mirror and 3LE. The persistent oscillation is called
oscillating bound state in a waveguide coupling to a giant atom with
multiple coupling points. Our study shows that the phenomenon of oscillating
bound states is also possible for a waveguide QED system with point-like
quantum emitters.

\begin{acknowledgments}
This work was supported by NSFC Grants No.11935006, No. 12075082, No. 12247105,
the science and technology innovation Program of Hunan Province (Grant No. 2020RC4047),
XJ2302001,and the Science $\And $ Technology Department of Hunan Provincial Program (2023ZJ1010).
\end{acknowledgments}


\begin{thebibliography}{99}
\bibitem{CookPRA35} R.J. Cook, and P.W. Milonni, Quanum theory of an atom near
partial reflecting wall, Phys. Rev. A 35, 5081 (1987).

\bibitem{Alberpra46} G. Alber, Photon wave packets and spontaneous decay in
a cavity, Phys. Rev. A 46, R5338 (1992).

\bibitem{Meystr56} E. V. Goldstein and P. Meystre, Dipole-dipole interaction
in optical cavities, Phys. Rev. A 56, 5135 (1997).

\bibitem{Zoller66} U. Dorner and P. Zoller, Laser-driven atoms in
half-cavities, Phys. Rev. A 66, 023816 (2002).

\bibitem{RMP95(23)} A.S. Sheremet, M.I. Petrov, I.V. Iorsh, A.V.
Poshakinskiy and A.N. Poddubny, Waveguide quantum electrodynamics:
Collective radiance and photon-photon correlations, Rev. Mod. Phys. 95,
015002 (2023).

\bibitem{ShenPRL95} J. T. Shen and S. Fan, Coherent Single Photon Transport
in a One-DimensionalWaveguide Coupled with Superconducting Quantum Bits,
Phys. Rev. Lett. 95, 213001 (2005).

\bibitem{LanPRL101} L. Zhou, Z. R. Gong, Y.-X. Liu, C. P. Sun, and F. Nori,
Controllable Scattering of a Single Photon inside a One-Dimensional
Resonator Waveguide,\ Phys. Rev. Lett. 101, 100501 (2008).

\bibitem{ZhouOE30} J. L. Tan, X. W. Xu, J. Lu, and L. Zhou, Few-photon
optical diode in a chiral waveguide, Opt. Express 30, 28696 (2022).

\bibitem{LanPRA78} L. Zhou, H. Dong, Y.-X. Liu, C. P. Sun, and F. Nori,
Quantum supercavity with atomic mirrors, Phys. Rev. A 78,
063827 (2008).

\bibitem{gongPRA78} Z. R. Gong, H. Ian, L. Zhou, and C. P. Sun, Controlling
quasibound states in a one-dimensional continuum through an
electromagnetically-induced-transparency mechanism, Phys. Rev. A 78, 053806
(2008).

\bibitem{DongPRA79} H. Dong, Z. R. Gong, H. Ian, L. Zhou, and C. P. Sun,
Intrinsic cavity QED and emergent quasinormal modes for a single photon,
Phys. Rev. A 79, 063847 (2009).

\bibitem{HoiPRL107} I.-C. Hoi, C. M. Wilson, G. Johansson, T. Palomaki, B.
Peropadre, and P. Delsing, Demonstration of a Single-Photon Router in the
Microwave Regime, Phys. Rev. Lett. 107, 073601 (2011).

\bibitem{lanPRL111} L. Zhou, L. P. Yang, Y. Li, and C. P. Sun, Quantum
Routing of Single Photons with a Cyclic Three-Level System Phys. Rev. Lett.
111, 103604 (2013).

\bibitem{LuPRA89} J. Lu, L. Zhou, L. M. Kuang, and F. Nori, Single-photon
router: Coherent control of multichannel scattering for single photons with
quantum interferences, Phys. Rev. A 89, 013805 (2014).

\bibitem{LuOE23} J. Lu, Z. H. Wang, and L. Zhou, T-shaped single-photon
router, Opt. Exp. 23, 22955 (2015)

\bibitem{routPRA94} C. Gonzalez-Ballestero, E. Moreno, F. J. Garcia-Vidal,
and A. Gonzalez-Tudela, Nonreciprocal few-photon routing schemes based on
chiral waveguide-emitter couplings, Phys. Rev. A 94, 063817 (2016).

\bibitem{WeiPRA89} C. H. Yan, Y. Li, H. D. Yuan, and L. F. Wei, Targeted
photonic routers with chiral photon-atom interactions, Phys. Rev. A 97,
023821 (2018).

\bibitem{AhumPRA99} M. Ahumada, P. A. Orellana, F. Dom\'{\i}guez-Adame, and
A. V. Malyshev, Tunable single-photon quantum router, Phys. Rev. A 99,
033827 (2019).

\bibitem{routPRR02} B. Poudyal and I.M. Mirza, Collective photon routing
improvement in a dissipative quantum emitter chain strongly coupled to a
chiral waveguide QED ladder, Phys. Rev. Res. 2, 043048 (2020)

\bibitem{LanPRA89} Z. H. Wang, L. Zhou, Y. Li, and C. P. Sun, Controllable
single-photon frequency converter via a one-dimensional waveguide, Phys.
Rev. A 89, 053813 (2014).

\bibitem{LiPRA104} L. Du and Y. Li, Single-photon frequency conversion via a
giant $\Lambda $-type atom, Phys. Rev. A 104, 023712 (2021).

\bibitem{YaAPL123} Y.-K. Luo, Ya Yang, Jing Lu, and Lan Zhou, Control of a
single-photon router via an extra cavity, Appl. Phys. Lett. 123, 211103
(2023).

\bibitem{YaAPL124} Y. Yang, J. Lu, and Lan Zhou, Controllable nonreciprocal
single-photon frequency converter via afour-level system, Appl. Phys. Lett.
124, 141102 (2024).

\bibitem{ShenPRL98} J. T. Shen and S. H. Fan, Strongly
Correlated Two-Photon Transport in a One-Dimensional Waveguide Coupled to a
Two-Level System, Phys. Rev. Lett. 98, 153003 (2007).

\bibitem{LiaoPRA82} J. Q. Liao and C. K. Law, Correlated two-photon
transport in a one-dimensional waveguide side-coupled to a nonlinear cavity,
Phys. Rev. A 82, 053836 (2010).

\bibitem{RoyPRL106} D. Roy, Two-Photon Scattering by a Driven Three-Level
Emitter in a One-Dimensional Waveguide and Electromagnetically Induced
Transparency, Phys. Rev. Lett. 106, 053601 (2011).

\bibitem{Zheng107} H. X. Zheng, D. J. Gauthier, and H. U. Baranger,
Cavity-Free Photon Blockade Induced by Many-Body Bound States, Phys. Rev.
Lett. 107, 223601 (2011).

\bibitem{ShiPRA84} T. Shi, S. H. Fan, and C. P. Sun, Two-photon transport in
a waveguide coupled to a cavity in a two-level system, Phys. Rev. A 84,
063803 (2011).

\bibitem{XuPRA90} X. W. Xu and Y. Li, Strongly correlated two-photon
transport in a one-dimensional waveguide coupled to a weakly nonlinear
cavity, Phys. Rev. A 90, 033832 (2014).

\bibitem{LanOE30} Y. L. Wang, Y. Yang, J. Lu, and L. Zhou, Photon transport
and interference of bound states in a one-dimensional waveguide, Opt. Exp.
30, 14048 (2022).

\bibitem{YaOE31} Y. Yang, J. Lu, and L. Zhou, Control of photon-photon
interaction via a cavity, Opt. Exp. 31, 39784 (2024).

\bibitem{ZhouPRA80} L. Zhou, S. Yang, Y.-x. Liu, C.P. Sun and F. Nori,
Quantum Zeno switch for single-photon coherent transport, Phys. Rev. A 80,
062109 (2009).

\bibitem{sunPRA100} L. Qiao and C.-P. Sun, Atom-photon bound states and
non-Markovian cooperative dynamics in coupled-resonator waveguides, Phys.
Rev. A 100, 063806 (2019).

\bibitem{WangPRA101} W. Zhao and Z.H. Wang, Single-photon scattering and
bound states in an atom-waveguide system with two or multiple coupling
points, Phys. Rev. A 101, 053855 (2020)

\bibitem{luOL49} Z.L. Lu, J. Li,, J. Lu, and L. Zhou, Controlling
atom-photon bound states in a coupled resonator array with a two-level
quantum emitter, Optics Letters, 49, 806 (2024).

\bibitem{LiNJP26} Jing Li, Jing Lu, Z R Gong and Lan Zhou, Tunable chiral
bound states in a dimer chain of coupled resonators, New J. Phys. 26, 033025
(2024).

\bibitem{SEPRA89} F. Lombardo, F. Ciccarello, and G. M. Palma, Photon
localization versus population trapping in a coupled-cavity array, Phys.
Rev. A 89, 053826 (2014).

\bibitem{SEPRA96} E. S\'{a}chez-Burillo, et al., Dynamical signatures of
bound states in waveguide QED, Phys. Rev. A 96, 023831 (2017).

\bibitem{ShenPRA87} M. Bradford and J. T. Shen, Spontaneous emission in
cavity QED with a terminated waveguide, Phys. Rev. A 87, 063830 (2013).

\bibitem{PRA842011} B. Peropadre, G. Romero, G. Johansson, C. M. Wilson, E.
Solano, and J. J. Garc\'{\i}a-Ripoll, Approaching perfect microwave
photodetection in circuit QED, Phys. Rev. A 84, 063834 (2011).

\bibitem{TTCKPRA87} T. Tufarelli, F. Ciccarello, and M. S. Kim, Dynamics of
spontaneous emission in a single-end photonic waveguide, Phys. Rev. A 87,
013820 (2013).

\bibitem{TTCKPRA90} T. Tufarelli, M. S. Kim, and F. Ciccarello,
Non-Markovianity of a quantum emitter in front of a mirror, Phys. Rev. A 90,
012113 (2014).

\bibitem{SongCTP69} H. X. Song, X. Q. Sun, J. Lu, and L. Zhou, Spatial
dependent spontaneous emission of an atom in a semi-infinite waveguide of
rectangular cross section, Commun. Theor. Phys. 69, 59 (2018).

\bibitem{NMPRA104} R. Trivedi and D. Malz, Optimal two-photon excitation of
bound states in non-Markovian waveguide QED, Phys. Rev. A 104, 013705 (2021).

\bibitem{PRA105Yi} L. Xin,, S. Xu, X. X. Yi, and H. Z. Shen, Tunable
non-Markovian dynamics with a three-level atom mediated by the classical
laser in a semi-infinite photonic waveguide, Phys. Rev. A 105, 053706 (2022).

\bibitem{RistPRA78} S. Rist, J. Eschner, M. Hennrich, and G. Morigi,
Photon-mediated interaction between two distant atoms, Phys. Rev. A 78,
013808 (2008).

\bibitem{HuQIP19} L. J. Hu, G. Y. Lu, J. Lu, and L. Zhou, Concurrence of two
identical atoms in a rectangular waveguide: Linear approximation with single
excitation, Quant. Info. Proc. 19, 81 (2020).
\end{thebibliography}
\end{document}